\documentclass{egpubl}
 \pdfoutput=1
\usepackage[T1]{fontenc}
\usepackage{dfadobe}  

\biberVersion
\BibtexOrBiblatex
\usepackage[backend=biber,bibstyle=EG,citestyle=alphabetic,backref=true]{biblatex} 
\electronicVersion
\PrintedOrElectronic

\ifpdf \usepackage[pdftex]{graphicx} \pdfcompresslevel=9
\else \usepackage[dvips]{graphicx} \fi

\usepackage{egweblnk} 
\usepackage{subfigure}

\title[Assessing Photorealism of Rendered Objects in Real-World Images: A Transparent and Reproducible User Study]%
      {Assessing Photorealism of Rendered Objects in Real-World Images: A Transparent and Reproducible User Study}

\author[S. Kluge \& O. Staadt]
{\parbox{\textwidth}{\centering S. Kluge\thanks{sven.kluge@uni-rostock.de}$^{1}$\orcid{0009-0008-3670-7098}
        and O. Staadt$^{1}$\orcid{0000-0002-3074-943X} 
        }
        \\
{\parbox{\textwidth}{\centering $^1$University of Rostock, Institute for Visual and Analytic Computing, Germany\\
    }
}
}
\begin{document}

\teaser{  
    \centering
    \subfigure[]{\includegraphics[width=0.19\textwidth]{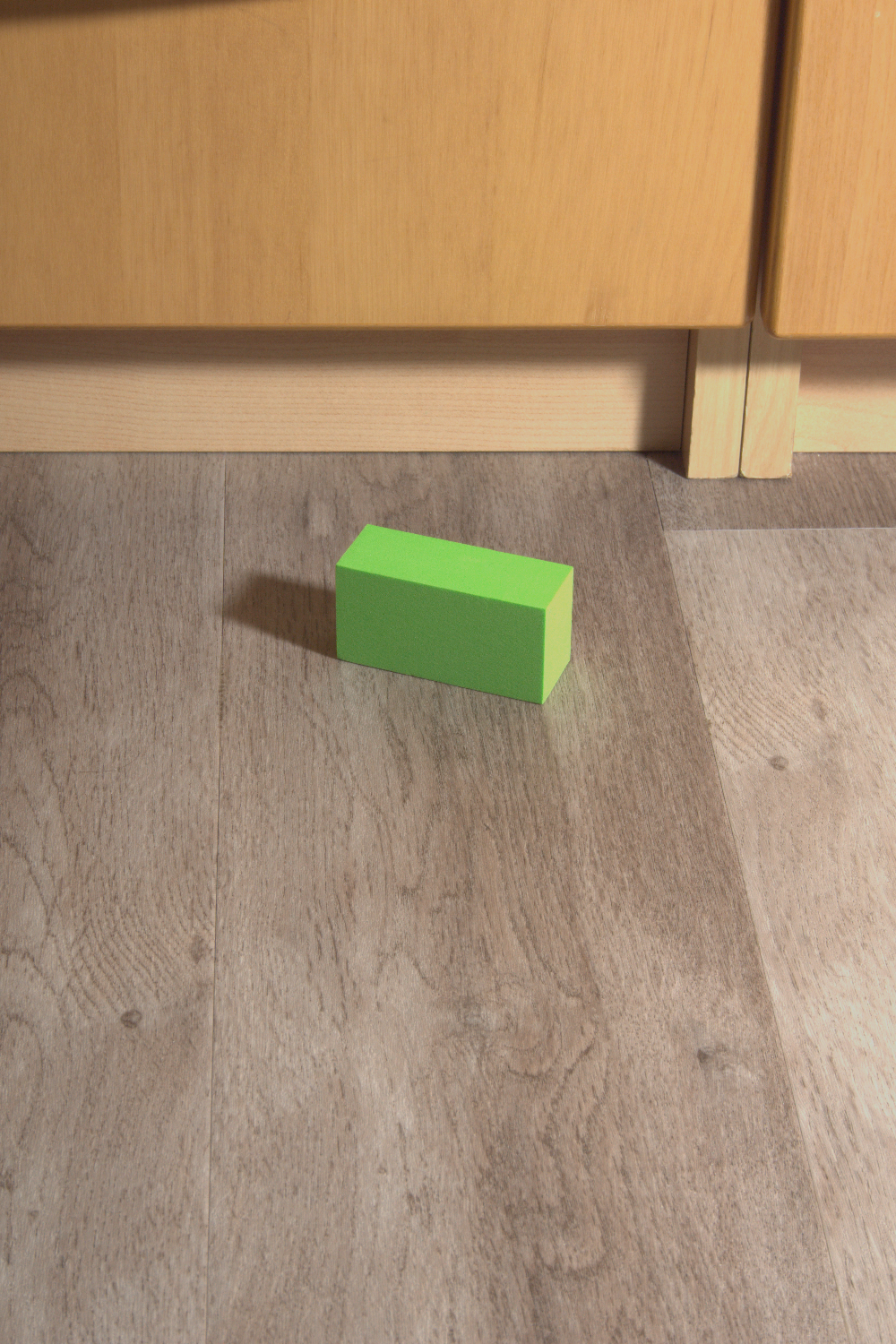}}
    \subfigure[]{\includegraphics[width=0.19\textwidth]{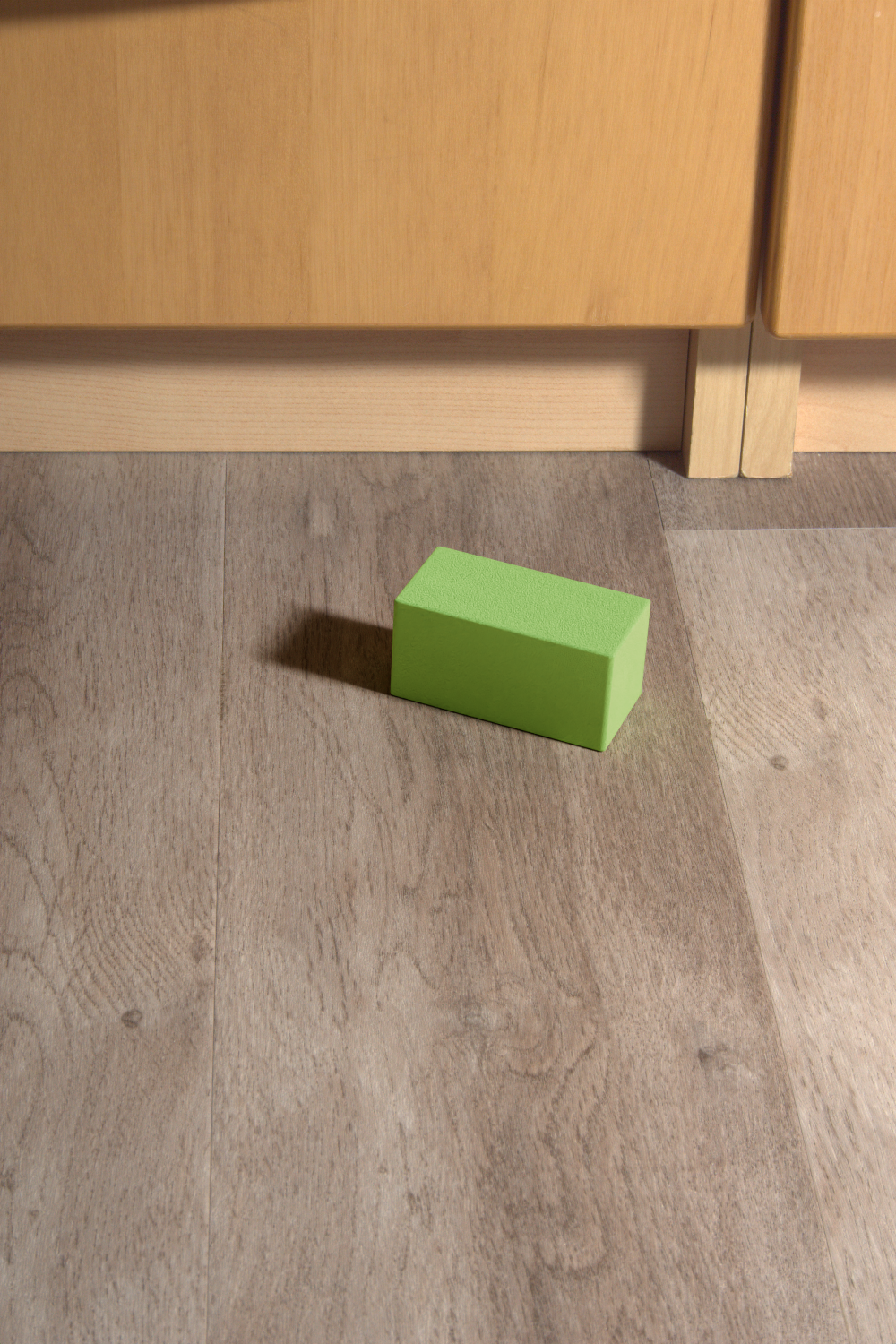}}
    \subfigure[]{\includegraphics[width=0.19\textwidth]{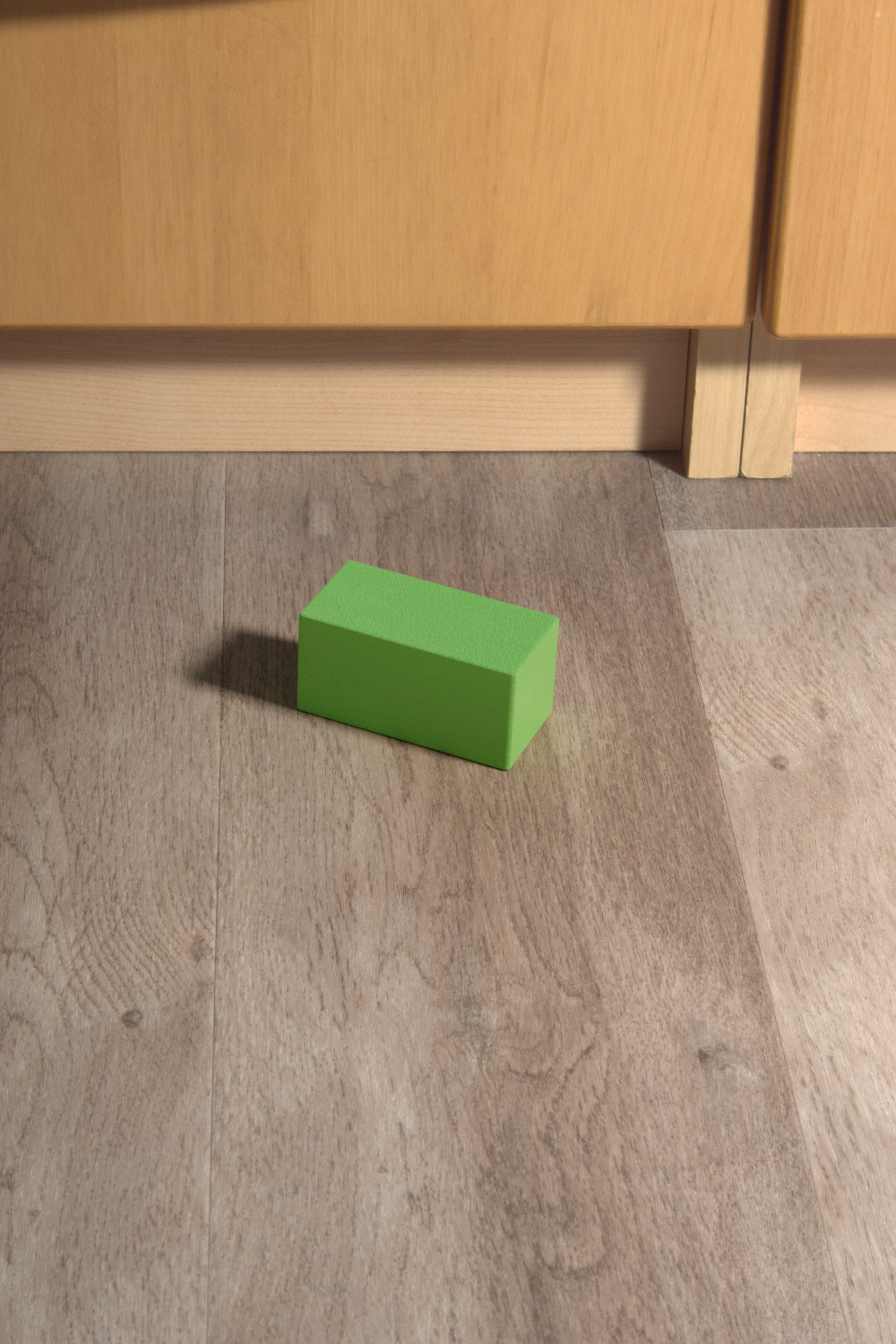}}
    \subfigure[]{\includegraphics[width=0.19\textwidth]{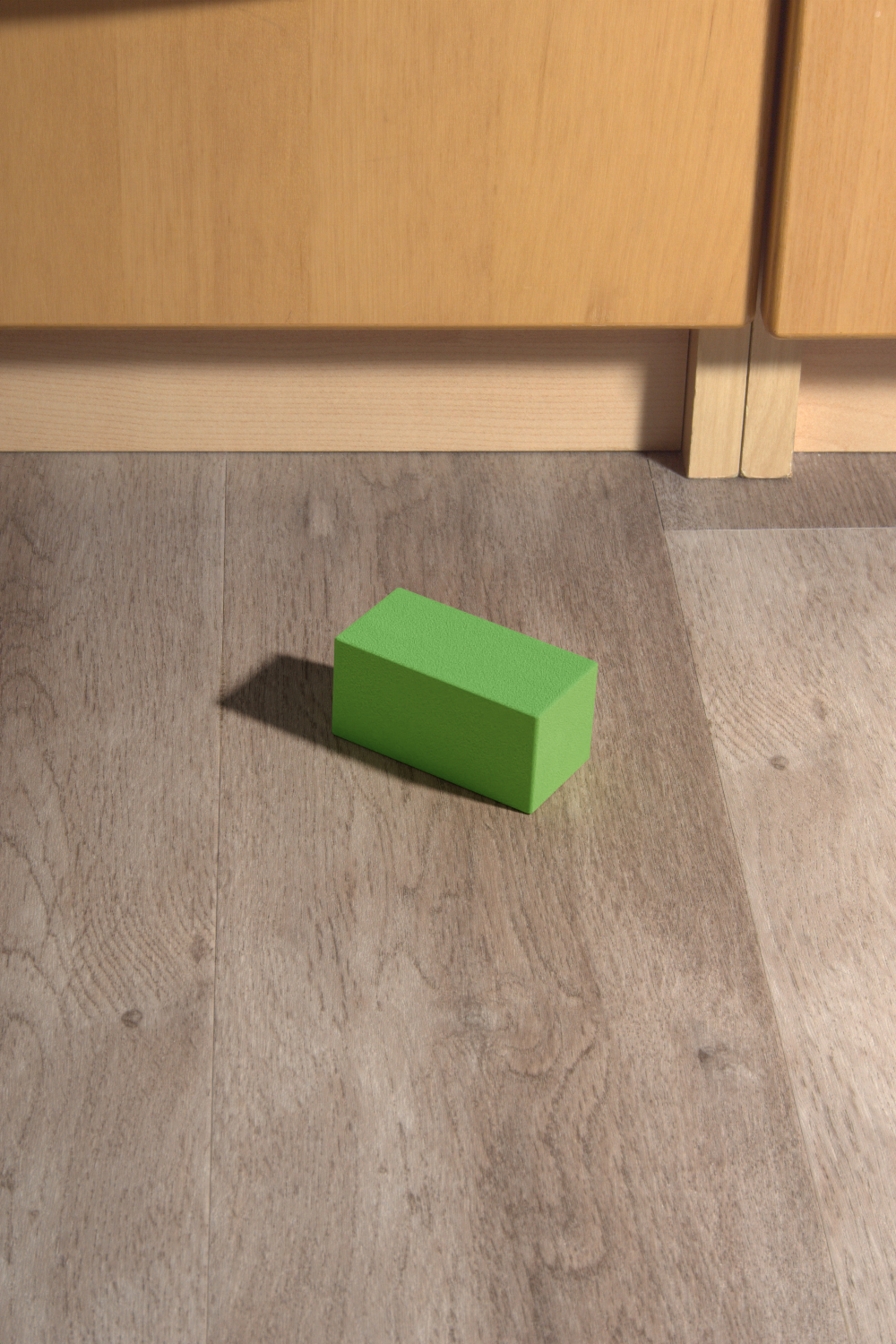}}
    \subfigure[]{\includegraphics[width=0.19\textwidth]{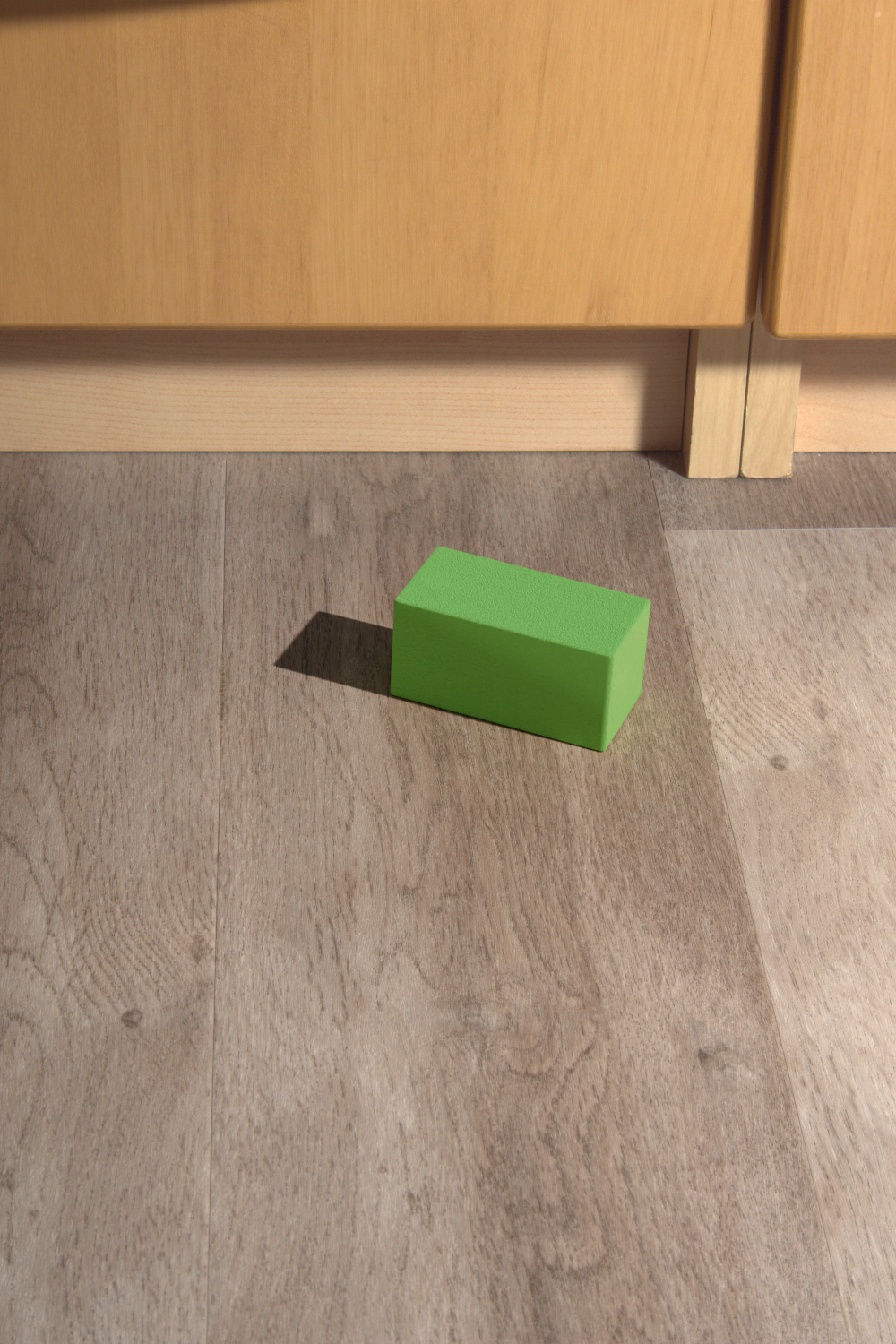}}
    \caption{Rendered objects composited into a real background; (a) reference (real) image, (b) object rendered with HDRI environment, (c-e) objects rendered with a directional light source and different shadow penumbra angles in order 10, 5, and 1 degree.}
    \label{fig:side-by-side}
}

\maketitle
%-------------------------------------------------------------------------
\begin{abstract}
    In an era where numerous studies claim to achieve almost photorealism with real-time automated environment capture, there is a need for assessments and reproducibility in this domain. This paper presents a transparent and reproducible user study aimed at evaluating the photorealism of real-world images composed with virtual rendered objects, that have been generated using classical environment capturing and rendering techniques. We adopted a two-alternative forced choice methodology to compare pairs of images created by integrating virtual objects into real photographs, following a classic pipeline. A control group with defined directional light parameters was included to validate the study's correctness. The findings revealed some insights, suggesting that observers experienced difficulties in differentiating between rendered and real objects. This work establishes the groundwork for future studies, aimed at enhancing the visual fidelity and realism of virtual objects in real-world environments. 
    
%-------------------------------------------------------------------------
%  ACM CCS 1998
%  (see https://www.acm.org/publications/computing-classification-system/1998)
% \begin{classification} % according to https://www.acm.org/publications/computing-classification-system/1998
% \CCScat{Computer Graphics}{I.3.3}{Picture/Image Generation}{Line and curve generation}
% \end{classification}
%-------------------------------------------------------------------------
%  ACM CCS 2012
%   (see https://www.acm.org/publications/class-2012)
%The tool at \url{http://dl.acm.org/ccs.cfm} can be used to generate
% CCS codes.
%Example:
\begin{CCSXML}
<ccs2012>
<concept>
<concept_id>10010147.10010371.10010372</concept_id>
<concept_desc>Computing methodologies~Rendering</concept_desc>
<concept_significance>300</concept_significance>
</concept>
<concept>
<concept_id>10003120.10003145.10011769</concept_id>
<concept_desc>Human-centered computing~Empirical studies in visualization</concept_desc>
<concept_significance>300</concept_significance>
</concept>
<concept>
<concept_id>10003120.10003121.10003122.10003334</concept_id>
<concept_desc>Human-centered computing~User studies</concept_desc>
<concept_significance>300</concept_significance>
</concept>
</ccs2012>
\end{CCSXML}

\ccsdesc[300]{Human-centered computing~User studies}
\ccsdesc[300]{Human-centered computing~Empirical studies in visualization}
\ccsdesc[300]{Computing methodologies~Rendering}

\printccsdesc   
\end{abstract}  
%-------------------------------------------------------------------------
\section{Introduction}
Augmented Reality (AR) has emerged as a transformative technology, offering immersive and interactive experiences by augmenting the physical world with virtual content. While early AR systems focused on overlaying simplistic graphical elements, recent advancements have pushed the boundaries of realism, enabling virtual objects to seamlessly blend into the real environment. However, achieving photo-realistic rendering in real-time AR applications remains a considerable challenge.

One of the challenges is to accurately capture the environmental lighting to enhance the realism of rendering virtual objects. Recently, the scientific community has made significant strides in automating real-time environment capturing, with promising advancements towards achieving photorealism. Evaluations have typically involved objective comparisons to ground truth as well as small-scale user studies, shedding valuable light on the progress made in this domain.

Photorealism is a key factor in creating convincing and compelling AR experiences. It entails accurately simulating the physical characteristics of light, materials, and shadows, mimicking the behavior of real-world objects. The quest for photorealism has been driven by the desire to bridge the gap between the virtual and real worlds, enabling users to perceive virtual objects as indistinguishable from their physical counterparts. 

In this paper, we present a rendering pipeline that is designed to create realistic-looking static images. We build upon a foundation of classic rendering techniques and algorithms to address the problems associated with real-time photorealistic rendering. Our pipeline comprises multiple stages for capturing the environment, rendering, and compositing the final image.

To evaluate the effectiveness of our pipeline, we conducted an uncontrolled subjective online study, where participants compared images rendered using our system. The study aimed to assess the perceptual realism of the rendered images and gather valuable feedback from users. The results of the study provide insights into the validity of our pipeline and methodology.

The remainder of this paper is organized as follows: Section 2 provides an overview of related work in the fields of photo-realistic rendering and augmented reality. Section 3 presents the methodology and technical details of our rendering pipeline and also the design and execution of the subjective online study, including evaluation metrics. Section 4 presents the results and analysis of the study, showcasing the achieved realism through qualitative assessments. In section 5 we discuss our study and our findings. Finally, Section 6 concludes the paper by summarizing our contributions, discussing the implications of our findings, and outlining potential avenues for future research in real-time photo-realistic augmented reality.

In summary, this paper addresses the challenge of creating a reproducible study for photo-realistic rendering of virtual objects in real images, as they are being used for augmented reality. By utilizing a classic approach and explaining our methodology in detail, our study contributes to the transparency and reproducibility of research in the field. We believe that this is essential for advancing scientific understanding and building upon existing work.

\begin{figure*}[t]
 \centering % avoid the use of \begin{center}...\end{center} and use \centering instead (more compact)
 \includegraphics[width=0.9\textwidth]{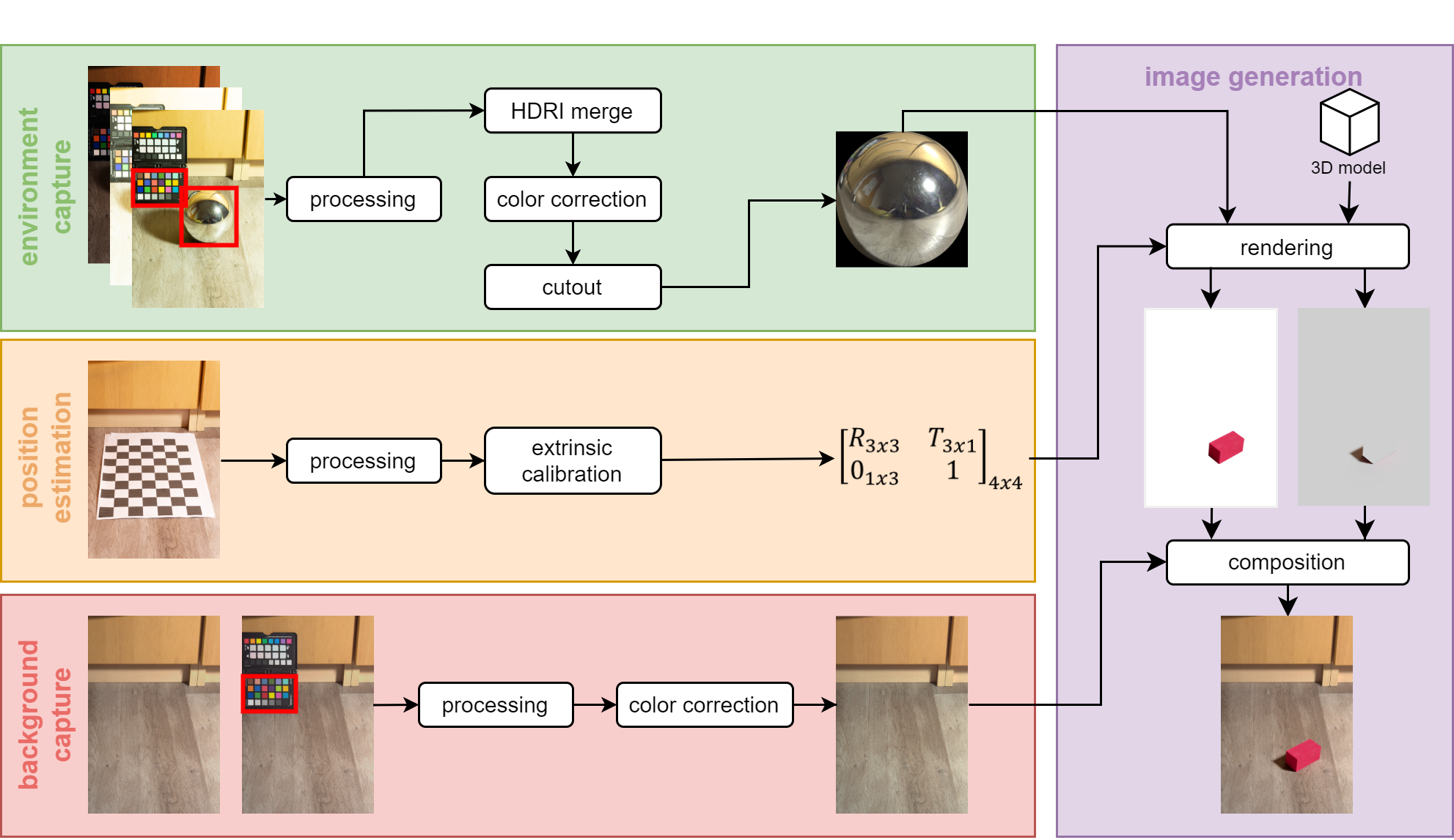}
 \caption{Rendering pipeline for photorealistic augmented reality with the three capturing stages environment capture, position estimation, and background capture. This information flows into the image generation stage, where the final image will be rendered and composed.}
 \label{fig:image_pipeline}
\end{figure*}

\section{Related Work}
Photorealistic AR rendering is an area of research that focuses on creating highly realistic virtual objects within real-world environments. Over the years, numerous researchers have contributed to the development of techniques and methodologies to achieve photorealistic rendering in AR. Debevec has made significant contributions to capturing and rendering realistic lighting environments. His work on light probes revolutionized the process of capturing environment lighting for photorealistic rendering. In~\cite{Debevec:1998}, Debevec presented a method for accurately integrating virtual objects into real-world scenes by capturing the illumination information using a set of high dynamic range (HDR) images known as light probes. The technique provided a foundation for subsequent research in photorealistic rendering and AR.

Gruber et al. \cite{Gruber:2014, Gruber:2015:ImagespaceIF} address the challenge of achieving photorealistic AR by computing accurate photometric registration. They propose an improved radiance transfer sampling approach that combines adaptive sampling in image and visibility space. By robustly caching radiance transfer, their method achieves real-time frame rates for dynamically changing AR scenes, enabling visually compelling and photorealistic experiences without invasive light probes.

Estimating environment lighting in an image using artificial intelligence (AI) techniques is an active research area that has gained significant attention in recent years. Numerous research studies have been published about the estimation of the environment light from a single image in indoor \cite{Gardner2017, Gardner2019} and outdoor environments \cite{HoldGeoffroy2019}. Some notable work has been done by Li. et al. \cite{Li2018, Li2019} who used intrinsic decomposition and neural inverse rendering to not only estimate the spatial varying lighting but also depth and albedo from a single image. This information can be used to improve the rendering of objects in AR. In order to devise a study aimed at comparing subjective visual realism, it is essential to understand the factors that can potentially impact perception. Various research has been conducted in the field, exploring the influence of various visual cues on the perceived realism of images.

Rademacher et al. \cite{Rademacher2001} conducted a study to quantify the perception of visual realism in images. They only used real images and investigated several visual factors, including shadow softness, surface smoothness, number of light sources, number of objects, and variety of object shapes. They found a borderline significance on the number of objects in the scene, and a significant correlation between the shadow penumbra angle and the perceived realism of the scene. They also found that the smoothness or roughness of the displayed object has a significant influence on the perception of realism. Another study conducted by Ferwerda et al. \cite{Ferwerda2010} investigated the perception of lighting errors. They defined the perceivable threshold for multiple parameters, including brightness errors, shading direction errors, shadow direction errors, and color temperature errors. Research by Lopez-Moreno \cite{LopezMoreno:2010} measured the accuracy of human vision in detecting lighting inconsistencies in images. Observers could only detect light divergence from the real light from 30-40 degrees in a real-world scenario.

\section{Image Generation Pipeline}

We developed an image generation pipeline comprising three stages to capture the environment and one stage for image generation (see Fig.~\ref{fig:image_pipeline}). While the pipeline itself contains classic algorithms and does not contain any novel scientific contribution, it is important for the reproducibility of our study that we define it as detailed as needed.

\subsection{Capturing}
To generate a composition of real images with rendered objects for augmented reality, we need to capture information about the camera and the real-world environment. In our pipeline, we captured the environment light, estimated the camera position, and captured the background for further image generation. For simplicity, we did not capture the environment geometry, because it was not necessary for occlusion handling in our test scene.

The environment can be captured with a classic light probe. We took multiple photos with varying exposure values (EV) from -2 to 20, with steps of 2 EV in between. This EV range allows us to capture most natural and artificial light sources. To merge the images into an HDR image, we used the hat weighting function, from Debevec et. al. \cite{Debevec:1997:RecoveringHD, Granados2010OptimalHR}. After the merge, the HDR image is color-corrected with a MacBeth ColorChecker chart \cite{mccamy:1976:color} that is visible in the photos and root-polynomial regression \cite{finlayson:2015}. After that, the light probe is cut out for usage in rendering.
The extrinsic camera parameters are estimated by taking a photo of a checkerboard pattern. The corner points between the black and white squares can be easily detected. Since we know, that all points are on a single plane, we can use a perspective-n-point pose estimation to calculate the extrinsic calibration \cite{marchand:2016}.
To make sure that the background and the HDRI light probe have the same color balance, we take two photos of the background. The first photo shows only the background, which is later used for image composition with the rendered object. The second photo shows the Macbeth ColorChecker chart, which is then used for color correction.

\subsection{Rendering}
After capturing the required environment information, the rendering process begins. Since we want to have high visual fidelity and realism with HDRI-based shadows, we are using a path tracer. The 3D model is rendered with the light probe as environment lighting. Because shadows are transparent, they have to be blended into the background. That is why we have to separate the rendering of the 3D model and the corresponding shadows. The resulting shadow image is blended onto the background, by using alpha-blending, and the rendered model is composited onto the resulting image.

\begin{figure}
    \centering
    \includegraphics[width=0.29\columnwidth]{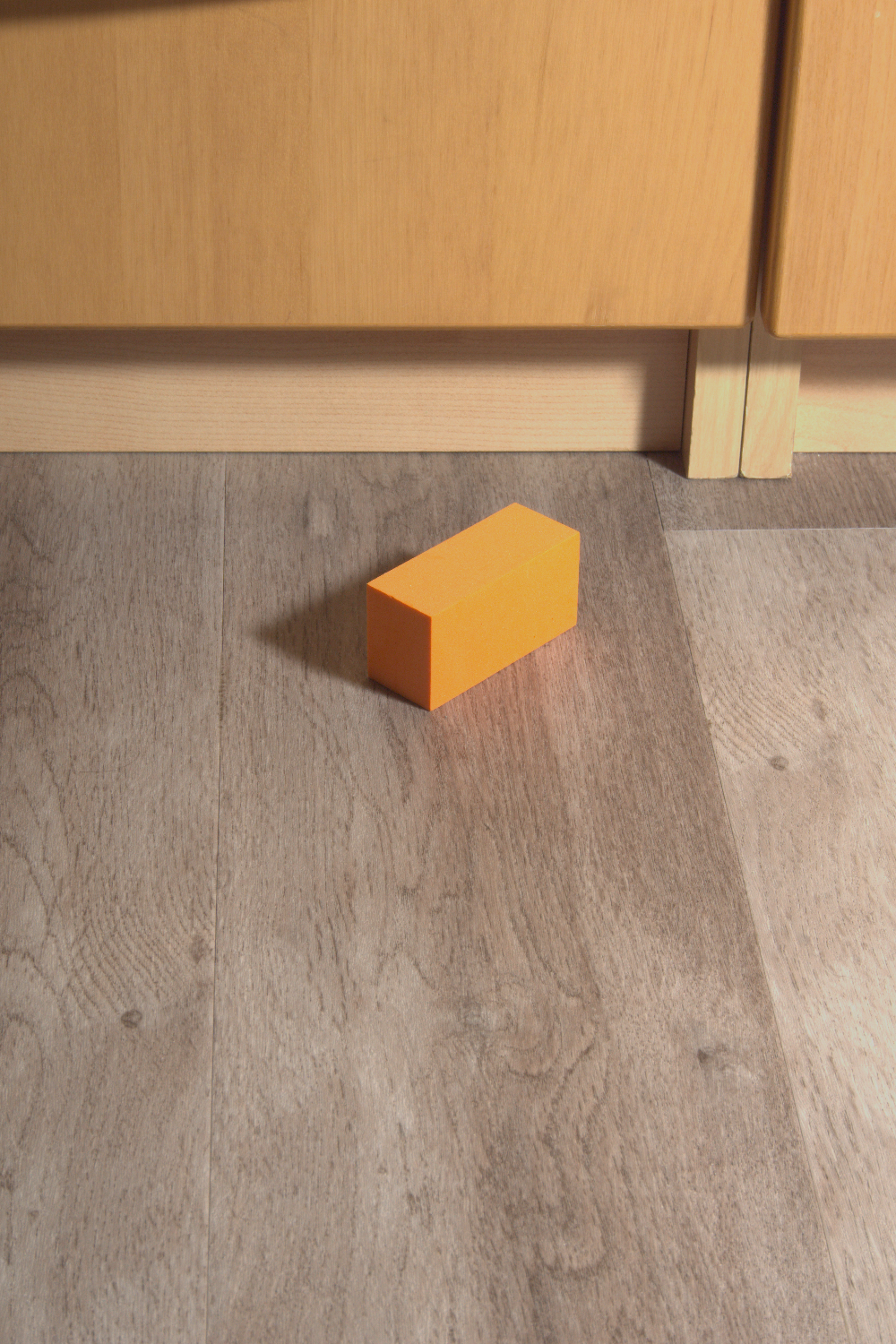}
    \includegraphics[width=0.29\columnwidth]{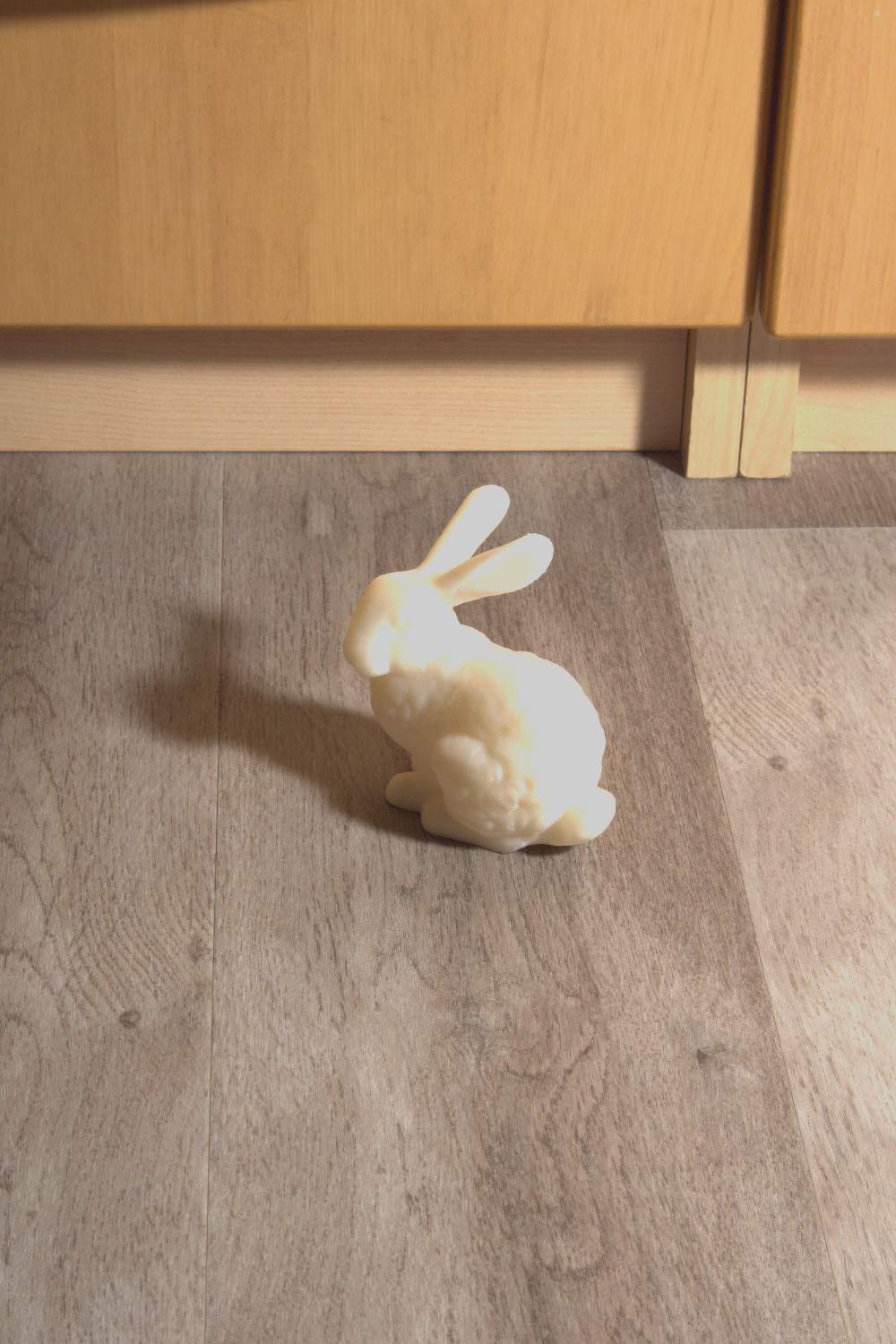}
    \includegraphics[width=0.29\columnwidth]{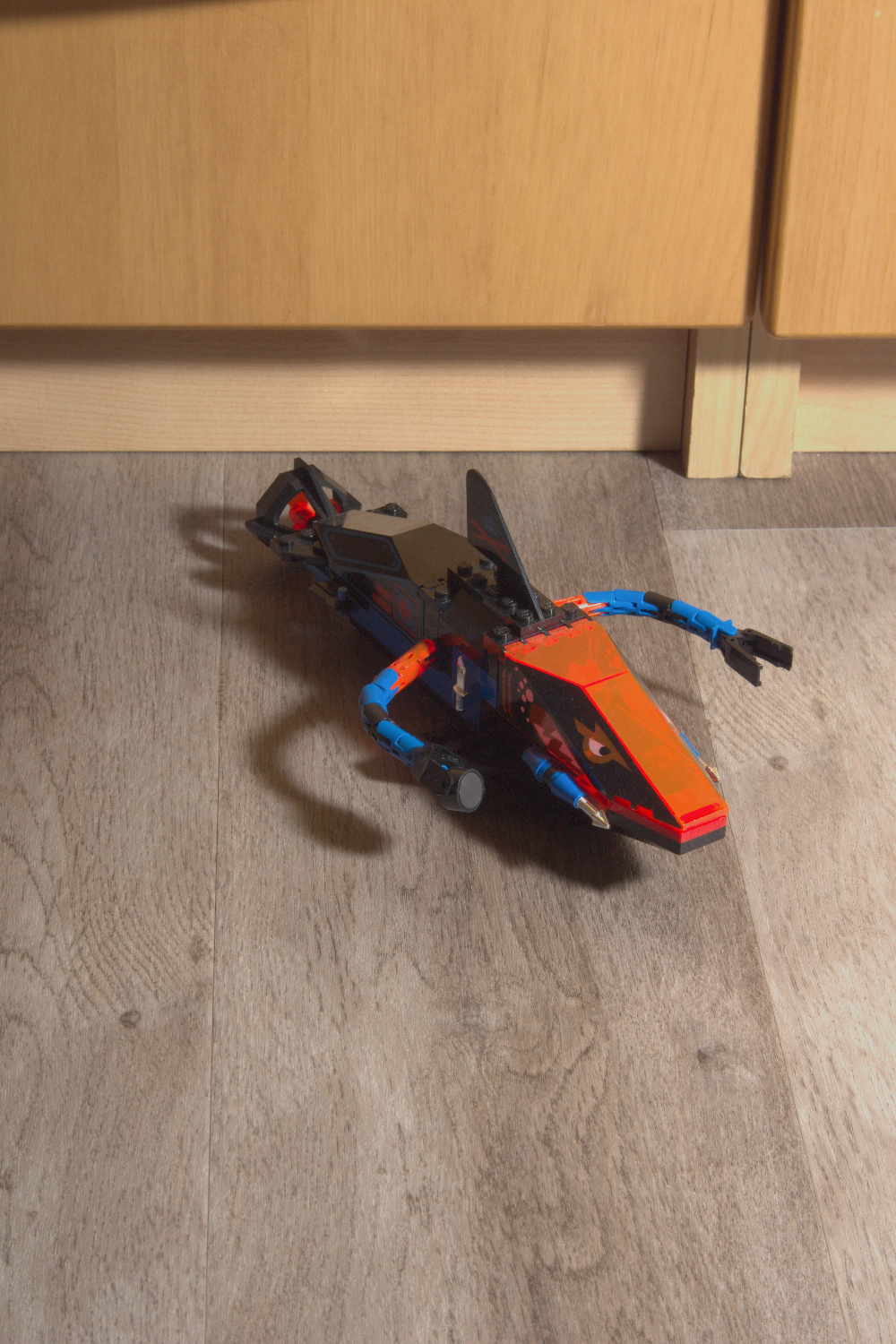}
    \includegraphics[width=0.29\columnwidth]{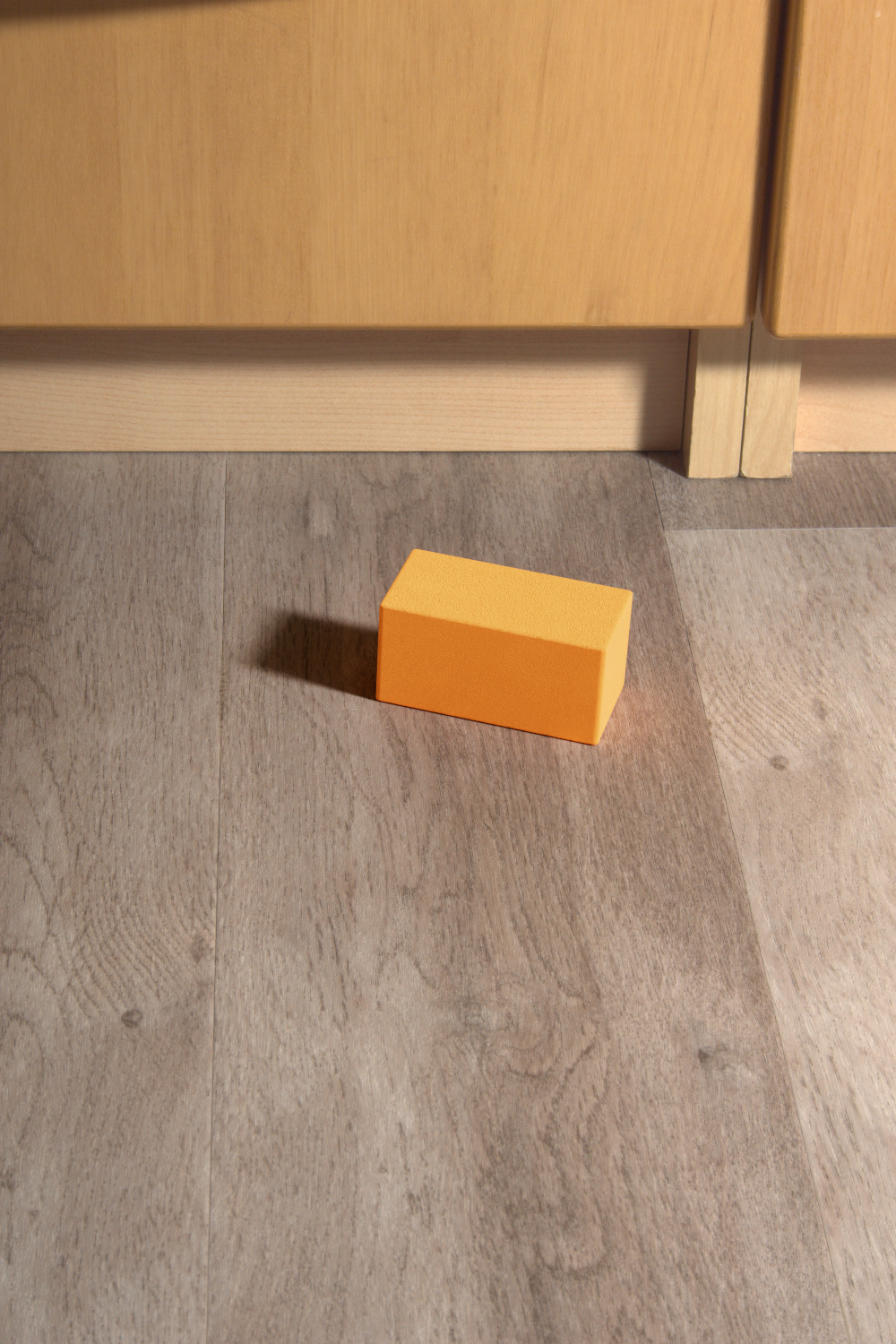}
    \includegraphics[width=0.29\columnwidth]{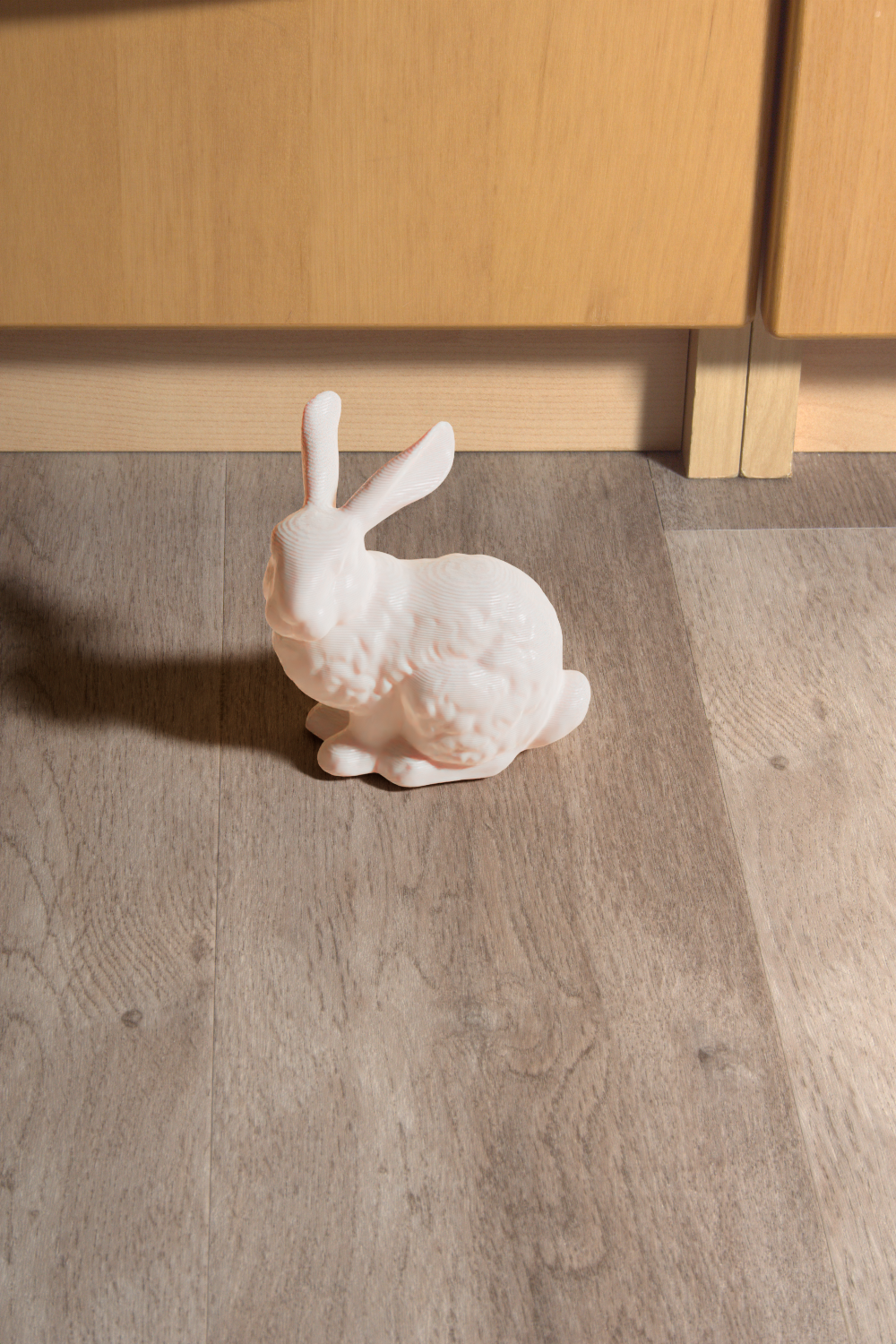}
    \includegraphics[width=0.29\columnwidth]{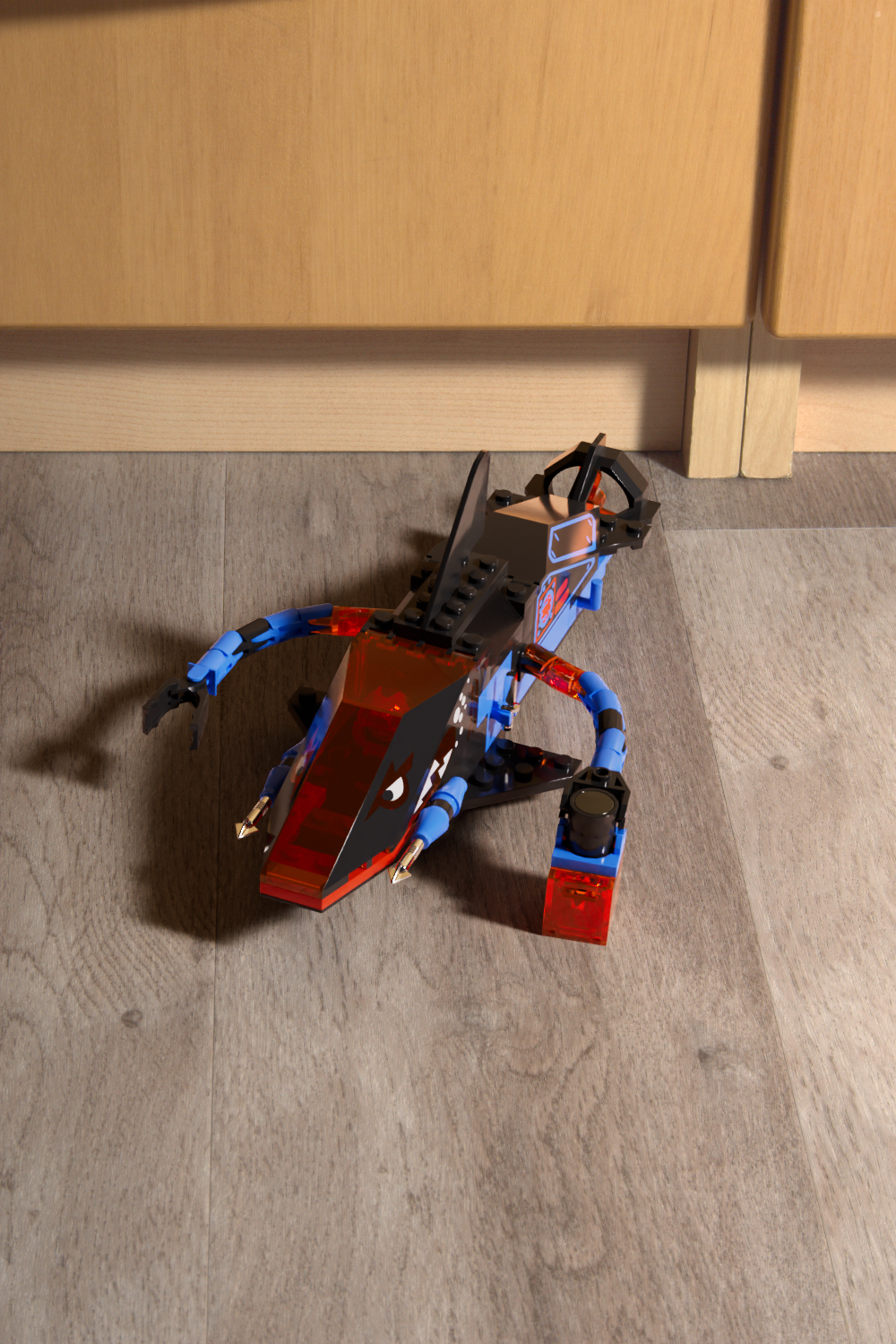}

    \caption{Samples of real models used in the study are in the upper row and their corresponding rendered counterparts, with HDRI environment mapping and path tracing. From left to right: foam toy building block, 3D printed bunny, Lego model.}
    \label{fig:my_label}
\end{figure}

\subsection{Stimuli Generation}

\subsubsection{Real Scene Setup}

The reference stimuli were created by taking photos of a real-world scene with a centrally placed object and defined lighting settings. We wanted to have different materials and shapes for our test models. We used four different types of small to medium-sized Lego models with glossy reflective surfaces, some of which have translucent parts. To include diffuse models, we used foam toy building blocks made from Ethylenvinylacetat in cylindrical and cubical shapes, in four different colors each. To include additional shapes, we added two 3D-printed Stanford bunny models to our study. Since the PLA printing material has some translucent properties, which are difficult to replicate in a rendering, we painted one of the models with a white primer. In total, we used 16 different models for our study.
To light the models, we used a light source with four fluorescent tubes. The lamp had a power of 27 W and a color temperature of 6500 K to simulate daylight. We positioned the lamp at a distance of 145 cm from the models, creating soft shadows. The models were lit from the front right, creating a well-visible soft shadow. The total size of our light source was 15 cm by 8 cm.

\subsubsection{Render Pipeline Implementation}
We implemented our offline rendering pipeline in Python, utilizing multiple packages and applications.
For raw processing of the images, we used the Python \textit{rawpy}\cite{rawpy} package, which is based on \textit{libraw}\cite{libraw}. We need to ensure that linear color space data is not altered in any undesirable way by \textit{libraw}, and validate every step of the process. It is important not to use any gamma correction or auto brightness setting. Highlights in raw images occur when sensor pixels are fully saturated. These pixel values are not reliable anymore and are greater than pure white would be. Especially for the HDR images, the brighter details should be included in images with a higher exposure value, so we clipped all highlight values. We used the AHD demosaicing algorithm to reconstruct the color image from our raw data~\cite{Hirakawa:2003:AdaptiveHD} and merged the resulting low dynamic range images to the HDR image using Python.
The position of the ColorChecker chart and the light probe in the image have been set manually for later color correction and for cutting out the light probe. Extrinsic camera calibration using perspective-n-point pose estimation \cite{marchand:2016} has been carried out using OpenCV.

To extract the colors from the ColorChecker chart, we manually set the rectangle, where the chart is located in the image. For processing, the image was converted into the CIE 1931 XYZ colorspace, and the color values from the checkerboard were extracted. The color value from the neutral gray field was used to calculate an initial exposure correction. We used the Python package \textit{colour} to calculate the root-polynomial regression for our color  correction with a degree of one and no root polynomial expansion.

For rendering, we employed the Cycles path tracer in Blender\cite{blender:2022:3.0}. To enhance rendering efficiency without compromising quality, we incorporated the \textit{Intel Open Image Denoise} API, which is an AI-based denoiser that reduces render times while maintaining details. This approach allowed us to achieve accurate renderings without the need for excessive sample counts. Cycles uses the Disney BRDF~\cite{Burley:2012:PhysicallyBasedSA} that offers flexibility in creating a wide range of materials with minimal parameters. However, it is important to note that the Disney BRDF is not necessarily physically accurate and does not guarantee perfect energy conservation.

To recreate the Lego models, we utilized the free BrickLink Studio 2.0\cite{bricklink} and employed the included materials, while other materials were handcrafted to resemble real-world counterparts. It is important to acknowledge that the goal of our study was not to achieve a perfect match between virtual and real-world materials but to compare their perceived realism and find a baseline for future studies and research. Because of their simple shape, the foam cuboids and foam cylinders have been measured and modeled as virtual objects. The respective materials have been visually approximated and created by hand.

\subsection{Experimental Design}

\begin{figure}[tb]
 \centering 
 \includegraphics[width=\columnwidth]{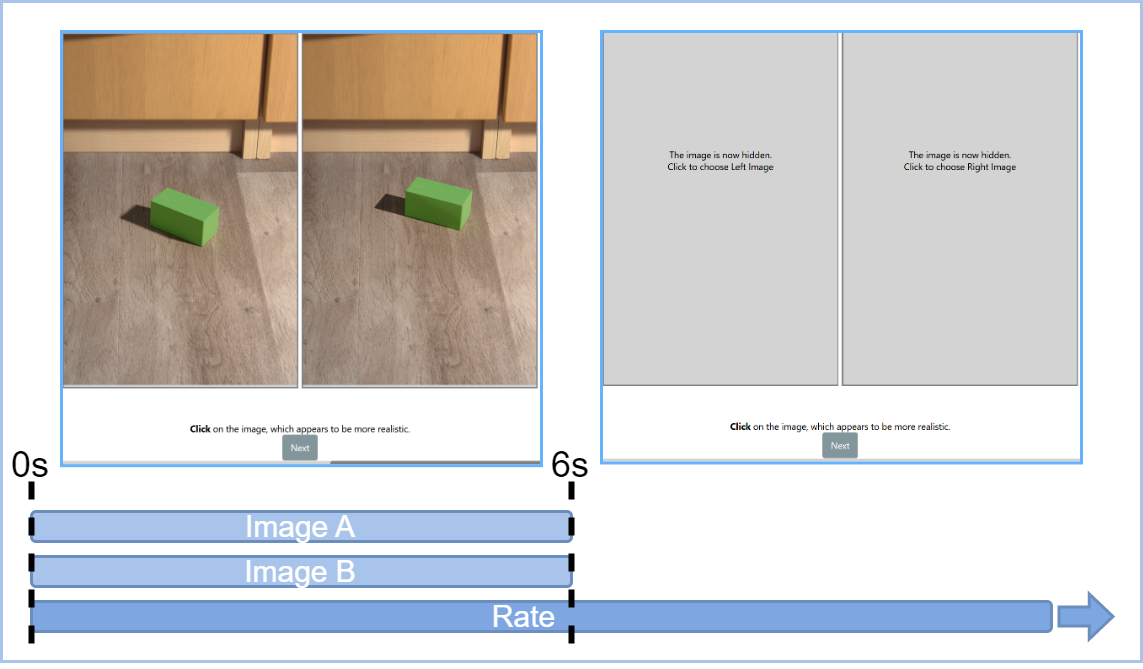}
 \caption{Representation of the two stages of a comparison task. Two images are shown for 6 seconds and then the images will be hidden. The observer may rate the images at any time and have additional time after the images are hidden.}
 \label{fig:study_design}
\end{figure}

To verify the realism of our images, we conducted an uncontrolled study using an online questionnaire based on the standard \cite{Rec.ITURBT.50014.2019} for image quality comparison. An uncontrolled study allows us to capture the natural variation present in real-world situations. Specifically, different display and environmental factors may change the perception of displayed images. However, these types of studies may also suffer from validity concerns because we cannot confidently attribute changes in participants' responses solely to the change in stimuli. So we must take extra measures to validate the responses and study results.

\subsubsection{Control Group}
The data for the study consisted of reference images and corresponding images with the virtual object, rendered using a captured HDRI environment map and path tracing. Since a small pilot study showed, that there was a high chance that observers would be able to reliably distinguish between different levels of realism, we included a control group. The purpose of the control group was to validate the study by having an outcome that was already known in advance. Previous studies have shown that the angle of the shadow penumbra influences perceived realism \cite{Rademacher2001}. The specific subjective perception of the shadow penumbra is influenced by the lighting and the scene. To ensure a measurable difference in subjective perception, we selected three additional groups with varying degrees of shadow penumbra.

The only way to create well-defined shadow penumbras in rendering is by using an explicit light source, such as a directional light. Therefore, we adjusted our rendering pipeline by placing a directional light that mimicked the position and characteristics of the real light source. The light temperature and strength were set according to the real light source.

The control groups were labeled as \textit{dir10}, \textit{dir5}, and \textit{dir1} to represent their corresponding degrees of shadow penumbra.

\subsubsection{Reduced Comparison}
To efficiently compare images with different degrees of realism among a larger number of groups, we employed a reduced comparison procedure. Creating a full comparison matrix for a single model, where there are $n$ different groups to compare, would require a total of $0.5(n*(n-1))$ comparisons. However, by utilizing an efficient sorting algorithm, the number of comparisons can be reduced to $n \log(n)$.

The concept behind reduced comparison is to compare only similar stimuli with each other, allowing observers to sort the stimuli using a sorting algorithm. The decrease in accuracy is negligible, as it enables us to collect more data within the same time frame \cite{Silverstein:2001:reduced_comparison}. To implement the reduced comparison, we utilized a self-balancing tree that maintained a shallow hierarchy. To evaluate the data, we can directly utilize the ordering from the sorting algorithm and assign a rank to each position in the order.

\subsubsection{Experimental Procedure}
The goal of this study is to evaluate the subjective realism of the generated images. For our study, we used a \textit{forced-choice with hidden reference} method. This method is widely used in psychological studies for its ease of use and accuracy and is also one of the recommendations for image quality comparisons \cite{Rec.ITURBT.50014.2019}. 

The participants were invited with a link to the online study, which they could open in their own browser on their own desktop computers.
Prior to the task, the participants were presented with an introduction about the topic of the study, followed by an interactive tutorial explaining the task. The participants were presented with two images side-by-side, from which they should choose the image that they perceived as being more realistic. To reduce the overall experiment time, the images began fading away after six seconds and were hidden (see Fig.~\ref{fig:study_design}). The remaining time was shown at the bottom of the screen as an unobtrusive progress bar. After the images were hidden the participants could take as long as they like to make their choice. After all the images were compared and rated, the participants were presented with a questionnaire about the task. This was intended to obtain additional feedback and to confirm that the participants had understood the task. At the end of the questionnaire, the participants  had the possibility to send textual feedback.

\begin{figure}[t]
    \centering
    \includegraphics[scale=0.9]{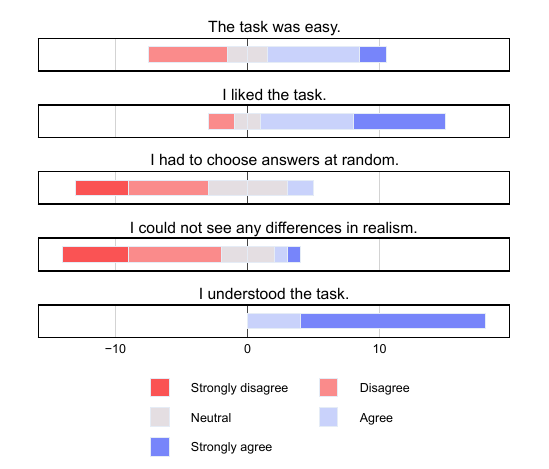}
    \caption{Results of the questionnaire that was conducted after the study, with five questions about the task. The results show, that observers thought that they understood the task. Many of the observers liked the task or were okay with it. Many observers agreed that the tasks were easy, but some found the task tedious or hard. Many observers felt that they had taken an informed choice on the tasks and didn't need to choose random answers because they could not see any differences.}
    \label{fig:result_likert}
\end{figure}

\section{Results}
In total 18 participants took part in the study, including university students and employees as well as outside participants. 
The statistical analysis was performed using Python including the packages SciPy \cite{SciPy-NMeth:2020}, statsmodels \cite{Seabold:2010}, and Pingouin \cite{Vallat:2018}. These packages provide robust functionalities for conducting statistical analyses.

\begin{figure}
    \centering
    \includegraphics[scale=1.0]{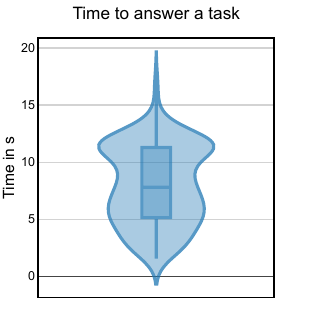}
    \caption{Violin plot with embedded box plot that visualizes how much time the observers took to answer comparison tasks. There are two accumulations of timings at around 6 s and 11 s. This resulted because the observers needed some more time to get accustomed to the task and got faster when they understood, that they could make an earlier choice.}
    \label{fig:timings_boxplot}
\end{figure}

\subsection{Screening Observers}
Before analyzing the results, it is necessary to filter out observers who may not have understood the task or randomly selected answers. The ITU-R Bt.500-14 Annex 1 \cite{Rec.ITURBT.50014.2019} provides guidelines for rejecting outliers. For each observer, their scores are compared to the mean value of all scores. Scores above the mean value plus the standard deviation times $\sqrt{20}$ and scores below the mean value minus the standard deviation times $\sqrt{20}$ are counted. If the sum of these counts, scaled by the number of observations made by the observer, exceeds the threshold of 0.05, the data is considered an outlier and rejected. Additionally, if the absolute difference of the counts divided by the sum of the counts is below the threshold of 0.3, the data from that observer is also rejected as an outlier. Based on these criteria, no observer was excluded from the analysis.

Furthermore, a plausibility check was conducted. One observer was excluded because they consistently answered in under 300ms. However, another observer who took a short break of approximately 30 seconds between tasks was included in the analysis as their data passed the initial screening procedure.

\subsection{Task Questionnaire}

The results depicted in Fig.~\ref{fig:result_likert} demonstrate that all observers felt confident in their understanding of the task. This suggests that our interactive tutorial was effective in providing clarity. The majority of participants perceived the task as easy, while only a few found it slightly difficult. This difficulty may be attributed to the subtle distinctions in realism that were challenging to discern. It aligns with the finding that some observers encountered difficulty in perceiving differences in realism and resorted to random answers. Despite these challenges, overall, the majority of participants expressed satisfaction with the task.

\subsection{Timings}

As mentioned previously, one observer took a short break between answering tasks. To ensure the accuracy of our timing analysis and the scaling of our visualization in Fig.~\ref{fig:timings_boxplot}, we applied MAD outlier removal with a cutoff value of 3.5 \cite{MAD-LEYS:2013} to remove excessively long answer times.

On average, participants required approximately 8.1 s to complete the task. Half of the tasks were answered within the time range of 5.1 s to 11.3 s. The plot reveals two distinct accumulations of task timings. The first accumulation occurs around 6 s when the images are being hidden, while the second accumulation occurs around 11.5 s. These differences can be attributed to observers taking their time to answer the initial tasks after the tutorial, gradually becoming more familiar and subsequently answering more swiftly. Conducting additional trial runs before the actual tasks may have resulted in more consistent timings. However, we found no evidence to suggest that these timing differences influenced the observers' choices regarding subjective realism.

\begin{figure}
    \centering
    \includegraphics[scale=1.0]{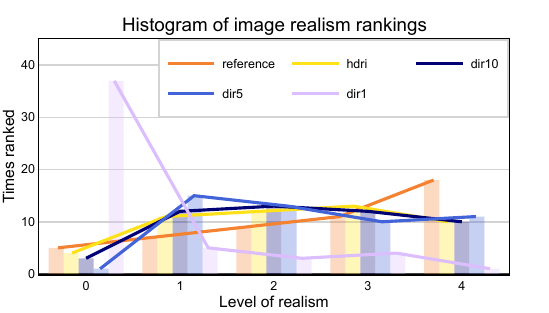}
    \caption{Histogram of how often a group ranked in a level of realism for the five groups used in the study. Ranked from least realism with level 0 to highest realism with level 4. \textit{reference} - real photos; \textit{hdri} - with HDRI environment map rendered images; \textit{dir10, dir5, dir1} - with directional light rendered images with shadow penumbra angles of 10, 5, and 1 degree.}
    \label{fig:results_realism}
\end{figure}

\subsection{Subjective Realism}

To identify any trends and patterns regarding the perceived realism among different groups, we plotted the histogram of rankings for each group as shown in Fig.~\ref{fig:results_realism}. Notably, the group \textit{dir1}, which featured the hardest shadows and the lowest physical accuracy, was predominantly associated with the lowest level of realism. Conversely, the \textit{reference} group stood out as having the highest level of perceived realism.

To conduct a statistical analysis, we initially performed a Kruskal-Wallis test \cite{kruskal:1952} on the rankings of the different groups to determine if the samples originated from the same distribution. The test yielded a significant difference between one or more groups (ddof=4, H=68.186, p$<$0.001). Subsequently, we conducted pairwise Wilcoxon rank-sum tests \cite{wilcoxon:1945} for further analysis. Given that this study is preliminary, we decided to report the results without any p-value adjustment, acknowledging the increased possibility of type I errors. Consequently, any significant findings are subject to a higher risk of being false positives and may require further investigation for confirmation.

The group \textit{dir1} exhibited a statistically significant difference with a very large effect size compared to all other groups: \textit{reference} (U=2,168.5, p=0.000, d=1.685), \textit{hdri} (U=2,136.5, p=0.000, d=1.508), \textit{dir10} (U=2,154, p=0.000, d=1.532), and \textit{dir5} (U=309, p=0.000, d=1.572). No other statistically significant differences were observed.

The \textit{dir1} group emerged as the group with the lowest perceived realism, which aligns with the findings of Rademacher et al. \cite{Rademacher2001}. Interestingly, no significant differences were observed between the \textit{dir5}, \textit{dir10}, and \textit{hdri} groups. The realism of the images from these groups appeared to be sufficiently similar to one another, making them indistinguishable. In future studies, it may be advisable to exclude non-significant control groups.

Although there was no statistical significance observed for the real images, in practice, observers found the \textit{reference} images to be more realistic compared to the \textit{hdri} group in 65\% of all direct comparisons. However, it is important to interpret this value with caution due to the use of the reduced comparison method. It is possible that in some cases, the absence of certain direct comparisons influenced this result.

After the study, we provided an opportunity for participants to provide written feedback, and six out of the 18 participants took advantage of this option. Most of the comments centered around the decision-making process regarding the realism of the images, which was based on the softness of the shadows. We believe this observation primarily corresponds to the presence of hard shadows in the \textit{dir1} group, as it was most apparent in that group. The noticeable differences in shadow penumbra in the control group may have encouraged participants to specifically compare the images based on the shadows. However, since the shadow of the \textit{hdri} group closely resembled the real shadow, we do not believe this influenced the results of our main comparison between the \textit{reference} and \textit{hdri} images. Further investigation may be necessary to explore this aspect in more detail.

\section{Discussion}

Considering that the virtual materials used in the study were all handcrafted, we suspect that some materials were easier to distinguish as real or rendered. However, due to the limited number of samples available for each material, we are unable to statistically analyze this issue. Further research could be conducted using different types of materials, including those that are handcrafted, automatically generated, or physically measured. Moreover, since our images already encompass a wide range of variations, we only employed a single background with the same lighting environment for all the images. It is important to note that other backgrounds and environments may significantly impact the perception of realism, and exploring these factors in future studies could yield valuable insights.

In comparing the \textit{reference} images to the \textit{hdri} images, we observed that the shadows in the rendered images appeared darker. We attribute this to the absence of a cabinet behind the object in the rendered images, which would reflect light from the back and result in a brighter shadow in real-world scenarios. Since we did not capture the geometry of the scene, our rendered images lack this effect, causing the shadows to appear darker. To test this hypothesis, we manually placed a virtual wall behind the object, resulting in a brighter shadow in our renderings. At present, we do not have a clear understanding of the extent to which shadow brightness and color influence the perception of realism. Further research is needed to investigate this aspect more comprehensively.

As we already suspected, there was no significant difference between the real images and the \textit{hdri} group. The control group showed the validity of the study execution. The control groups \textit{dir10} and \textit{dir5} have similar results, so the number of control groups could have been reduced to only the \textit{dir5} group. However, this is only true for our chosen background scene. There are not enough scientific studies to ensure a general threshold value between 5 and 1-degree shadow penumbra. For other scenes, the threshold value for the detection of real and virtual objects may be different.

\section{Conclusion and Future Work}

In this study, we have established a baseline for our rendering pipeline, demonstrating that our materials and setup enable us to generate images that are extremely difficult to distinguish from real images. The inclusion of a control group further validated the effectiveness of our study. 

Moving forward, there are several avenues for future research. One important direction is to investigate the key factors that contribute to achieving photorealistic augmented reality. By continually comparing any modifications to our pipeline against the results of this preliminary baseline study, we can gain valuable insights into the impact of various factors on perceived realism. Of particular interest is the adaptation of our pipeline to real-time AR scenarios, involving see-through head-mounted displays or other AR devices. This presents an exciting opportunity to explore the challenges and possibilities of rendering photorealistic imagery in a dynamic and interactive AR environment. Conducting a similar user study in the context of real-time AR and comparing the results to those obtained in this study would provide valuable insights and help assess the transferability of our findings.

In conclusion, our study lays the foundation for future research in achieving photorealistic AR experiences. By refining our rendering pipeline and investigating additional factors, we aim to further enhance the realism of virtual content in AR applications.

%-------------------------------------------------------------------------
% bibtex
%\bibliographystyle{eg-alpha-doi}  
%\bibliography{bibliography}        

% biblatex with biber
\printbibliography                

@article{marchand:2016,
  TITLE = {{Pose Estimation for Augmented Reality: A Hands-On Survey}},
  AUTHOR = {Marchand, Eric and Uchiyama, Hideaki and Spindler, Fabien},
  URL = {https://inria.hal.science/hal-01246370},
  JOURNAL = {{IEEE Transactions on Visualization and Computer Graphics}},
  PUBLISHER = {{Institute of Electrical and Electronics Engineers}},
  VOLUME = {22},
  NUMBER = {12},
  PAGES = {2633 - 2651},
  YEAR = {2016},
  MONTH = Dec,
  DOI = {10.1109/TVCG.2015.2513408},
  HAL_ID = {hal-01246370},
  HAL_VERSION = {v1},
}

@article{finlayson:2015,
author = {Finlayson, Graham and Mackiewicz, Michal and Hurlbert, Anya},
year = {2015},
month = {05},
pages = {1460-1470},
title = {Color Correction Using Root-Polynomial Regression},
volume = {24},
journal = {IEEE transactions on image processing : a publication of the IEEE Signal Processing Society},
doi = {10.1109/TIP.2015.2405336 }
}

@misc{Rademacher2001,
 author = {Rademacher, Paul and Lengyel, Jed and Cutrell, Edward and Whitted, Turner},
 date = {2001},
 title = {Measuring the Perception of Visual Realism in Images},
 publisher = {{The Eurographics Association}},
 doi = {10.2312/EGWR/EGWR01/235-248}
}

@misc{Rec.ITURBT.50014.2019,
 author = {{Rec. ITU-R BT.500-14}},
 date = {2019},
 title = {Methodologies for the subjective assessment of the quality of television images}
}

@ARTICLE{Vallat:2018,
  title    = "Pingouin: statistics in Python",
  author   = "Vallat, Raphael",
  journal  = "The Journal of Open Source Software",
  volume   =  3,
  number   =  31,
  pages    = "1026",
  month    =  nov,
  year     =  2018
}

@inproceedings{Seabold:2010,
  title={statsmodels: Econometric and statistical modeling with python},
  author={Seabold, Skipper and Perktold, Josef},
  booktitle={9th Python in Science Conference},
  year={2010},
}

@ARTICLE{SciPy-NMeth:2020,
  author  = {Virtanen, Pauli and Gommers, Ralf and Oliphant, Travis E. and
            Haberland, Matt and Reddy, Tyler and Cournapeau, David and
            Burovski, Evgeni and Peterson, Pearu and Weckesser, Warren and
            Bright, Jonathan and {van der Walt}, St{\'e}fan J. and
            Brett, Matthew and Wilson, Joshua and Millman, K. Jarrod and
            Mayorov, Nikolay and Nelson, Andrew R. J. and Jones, Eric and
            Kern, Robert and Larson, Eric and Carey, C J and
            Polat, {\.I}lhan and Feng, Yu and Moore, Eric W. and
            {VanderPlas}, Jake and Laxalde, Denis and Perktold, Josef and
            Cimrman, Robert and Henriksen, Ian and Quintero, E. A. and
            Harris, Charles R. and Archibald, Anne M. and
            Ribeiro, Ant{\^o}nio H. and Pedregosa, Fabian and
            {van Mulbregt}, Paul and {SciPy 1.0 Contributors}},
  title   = {{{SciPy} 1.0: Fundamental Algorithms for Scientific
            Computing in Python}},
  journal = {Nature Methods},
  year    = {2020},
  volume  = {17},
  pages   = {261--272},
  adsurl  = {https://rdcu.be/b08Wh},
  doi     = {10.1038/s41592-019-0686-2   },
}

@article{MAD-LEYS:2013,
    title = {Detecting outliers: Do not use standard deviation around the mean, use absolute deviation around the median},
    journal = {Journal of Experimental Social Psychology},
    volume = {49},
    number = {4},
    pages = {764-766},
    year = {2013},
    issn = {0022-1031},
    doi = {https://doi.org/10.1016/j   .jesp.2013.03.013},
    url = {https://www.sciencedirect.com/science/article/pii/S0022103113000668},
    author = {Christophe Leys and Christophe Ley and Olivier Klein and Philippe Bernard and Laurent Licata}
}

@article{wilcoxon:1945,
 ISSN = {00994987},
 URL = {http://www.jstor.org/stable/3001968},
 author = {Frank Wilcoxon},
 journal = {Biometrics Bulletin},
 number = {6},
 pages = {80--83},
 publisher = {[International Biometric Society, Wiley]},
 title = {Individual Comparisons by Ranking Methods},
 urldate = {2023-06-10},
 volume = {1},
 year = {1945}
}

@article{kruskal:1952,
author = { William H.   Kruskal  and  W.   Allen   Wallis },
title = {Use of Ranks in One-Criterion Variance Analysis},
journal = {Journal of the American Statistical Association},
volume = {47},
number = {260},
pages = {583-621},
year  = {1952},
publisher = {Taylor & Francis},
doi = {10.1080/01621459.1952.10483441  },
URL = {https://www.tandfonline.com/doi/abs/10.1080/01621459.1952.10483441},
eprint = {https://www.tandfonline.com/doi/pdf/10.1080/01621459.1952.10483441}
}

@article{Debevec:1998,
  title={Rendering synthetic objects into real scenes: bridging traditional and image-based graphics with global illumination and high dynamic range photography},
  author={Paul E. Debevec},
  journal={Proceedings of the 25th annual conference on Computer graphics and interactive techniques},
  year={1998}
}

@inproceedings{Debevec:1997:RecoveringHD,
  title={Recovering high dynamic range radiance maps from photographs},
  author={Paul E. Debevec and Jitendra Malik},
  booktitle={International Conference on Computer Graphics and Interactive Techniques},
  year={1997}
}

@article{Granados2010OptimalHR,
  title={Optimal HDR reconstruction with linear digital cameras},
  author={Miguel Granados and Boris Ajdin and Michael Wand and Christian Theobalt and Hans-Peter Seidel and Hendrik P. A. Lensch},
  journal={2010 IEEE Computer Society Conference on Computer Vision and Pattern Recognition},
  year={2010},
  pages={215-222}
}

@inproceedings{LopezMoreno:2010,
 author = {Lopez-Moreno, Jorge and Sundstedt, Veronica and Sangorrin, Francisco and Gutierrez, Diego},
 title = {Measuring the perception of light inconsistencies},
 pages = {25},
 publisher = {{ACM Press}},
 isbn = {9781450302487  },
 editor = {Banks, Marty and Mania, Katerina and Gutierrez, Diego and Kearney, Joe and Spencer, Stephen N.},
 booktitle = {Proceedings of the 7th Symposium on Applied Perception in Graphics and Visualization - APGV '10},
 year = {2010},
 address = {New York, New York, USA},
 doi = {10.1145/1836248.1836252  }
}

@misc{Ferwerda2010,
 author = {Ferwerda, James A. and Selan, Jeremy and Pellacini, Fabio},
 date = {2010},
 title = {Perception of Lighting Errors in Image Compositing}
}

@inproceedings{Gruber:2014,
 author = {Gruber, Lukas and Langlotz, Tobias and Sen, Pradeep and Hoherer, Tobias and Schmalstieg, Dieter},
 title = {Efficient and robust radiance transfer for probeless photorealistic augmented reality},
 pages = {15--20},
 publisher = {IEEE},
 isbn = {978-1-4799-2871-2  },
 booktitle = {2014 IEEE Virtual Reality (VR)},
 year = {2014},
 doi = {10.1109/VR.2014.6802044  }
}

@article{Gruber:2015:ImagespaceIF,
  title={Image-space illumination for augmented reality in dynamic environments},
  author={Lukas Gruber and Jonathan Ventura and Dieter Schmalstieg},
  journal={2015 IEEE Virtual Reality (VR)},
  year={2015},
  pages={127-134}
}

@inproceedings{Burley:2012:PhysicallyBasedSA,
  title={Physically-Based Shading at Disney},
  author={Brent Burley},
  year={2012}
}

@article{Gardner2017,
 author = {Gardner, Marc-Andr{\'e} and Sunkavalli, Kalyan and Yumer, Ersin and Shen, Xiaohui and Gambaretto, Emiliano and Gagn{\'e}, Christian and Lalonde, Jean-fran{\c{c}}ois},
 year = {2017},
 title = {Learning to predict indoor illumination from a single image},
 pages = {1--14},
 volume = {36},
 number = {6},
 issn = {07300301},
 journal = {ACM Transactions on Graphics},
 doi = {10.1145/3130800.3130891 }
}

@misc{Gardner2019,
 author = {Gardner, Marc-Andr{\'e} and Hold-Geoffroy, Yannick and Sunkavalli, Kalyan and Gagn{\'e}, Christian and Lalonde, Jean-fran{\c{c}}ois},
 date = {2019},
 title = {Deep Parametric Indoor Lighting Estimation},
 url = {http://arxiv.org/pdf/1910.08812v1}
}

@incollection{Li2018,
 author = {Li, Zhengqi and Snavely, Noah},
 title = {CGIntrinsics: Better Intrinsic Image Decomposition Through Physically-Based Rendering},
 pages = {381--399},
 volume = {11207},
 publisher = {{Springer International Publishing}},
 isbn = {978-3-030-01218-2 },
 series = {Image Processing, Computer Vision, Pattern Recognition, and Graphics},
 editor = {Ferrari, Vittorio and Hebert, Martial and Sminchisescu, Cristian and Weiss, Yair},
 booktitle = {Computer Vision - ECCV 2018},
 year = {2018},
 address = {Cham},
 doi = {10.1007/978-3-030-01219-9 {\textunderscore }23}
}

@misc{Li2019,
 author = {Li, Zhengqin and Shafiei, Mohammad and Ramamoorthi, Ravi and Sunkavalli, Kalyan and Chandraker, Manmohan},
 date = {2019},
 title = {Inverse Rendering for Complex Indoor Scenes: Shape, Spatially-Varying  Lighting and SVBRDF from a Single Image},
 url = {http://arxiv.org/pdf/1905.02722v1}
}

@misc{HoldGeoffroy2019,
 author = {Hold-Geoffroy, Yannick and Athawale, Akshaya and Lalonde, Jean-fran{\c{c}}ois},
 date = {2019},
 title = {Deep Sky Modeling for Single Image Outdoor Lighting Estimation},
 url = {http://arxiv.org/pdf/1905.03897v1}
}

@article{mccamy:1976:color,
  title={A color-rendition chart},
  author={McCamy, Calvin S and Marcus, Harold and Davidson, James G and others},
  journal={J. App. Photog. Eng},
  volume={2},
  number={3},
  pages={95--99},
  year={1976}
}

@article{Silverstein:2001:reduced_comparison,
 author = {{D. Amnon Silverstein} and {Joyce E. Farrell}},
 year = {2001},
 title = {Efficient method for paired comparison},
 pages = {394--398},
 volume = {10},
 journal = {J. Electronic Imaging}
}

@article{Hirakawa:2003:AdaptiveHD,
  title={Adaptive homogeneity-directed demosaicing algorithm},
  author={Keigo Hirakawa and Thomas W. Parks},
  journal={Proceedings 2003 International Conference on Image Processing (Cat. No.03CH37429)},
  year={2003},
  volume={3},
  pages={III-669}
}

@misc{blender:2022:3.0,
  title = {Blender},
  author = {{Blender Foundation}},
  howpublished = {[Computer software]},
  year = {2022},
  note = {Version 3.0},
  url = {https://www.blender.org/}
}

@misc{rawpy,
  title = {rawpy},
  author = {{Maik Riechert}},
  howpublished = {[Computer software]},
  year = {2022},
  note = {Version 0.17.3},
  url = {https://pypi.org/project/rawpy/}
}

@misc{bricklink,
    title = {BrickLink Studio},
    author = {{BrickLink Corporation (2022)}},
    howpublished = {[computer software]},
    note = {Version 2.0},
    URL = {https://www.bricklink.com}
}

@misc{libraw,
    title = {libraw},
    author = {{LibRaw LLC (2023)}},
    howpublished = {[computer software]},
    note = {Version 0.20},
    url = {https://www.libraw.org/}
}

%-------------------------------------------------------------------------

\end{document}